\documentstyle[12pt,twoside,fleqn,espcrc1,psfig]{article}

\def\nuc#1#2{\relax\ifmmode{}^{#1}{\hbox{#2}}\else${}^{#1}$#2\fi}
\def\C12{\nuc{12}{C}}
\newcommand{\LamLam}{\mbox{$\Lambda$-$\Lambda$}}
\newcommand{\sLamLam}{{\scriptscriptstyle\Lambda\Lambda}}
\newcommand{\beqar}{\begin{eqnarray}}
\newcommand{\eeqar}{\end{eqnarray}}
\newcommand{\beq}{\begin{equation}}
\newcommand{\eeq}{\end{equation}}

\newcommand{\KK}{($K^-,K^+$)\ }
\newcommand{\reff}{\mbox{$r_{\rm eff}$}}

\title{ Can We Extract Lambda-Lambda Interaction \\
		from Two-Particle Momentum Correlation ? }

\author{Akira Ohnishi\address{%
	Division of Physics, Graduate School of Science,
                Hokkaido University, Sapporo 060-0810, Japan},
	Yuichi Hirata$^{\rm a}$, 
	Yasushi Nara\address{%
	Advanced Science Research Center,
		Japan Atomic Energy Research Institute,
		Japan}, 
	Shoji Shinmura\address{%
	Faculty of Engineering,
		Gifu University, 
		Japan}, 
	and Yoshinori Akaishi\address{%
	Institute of Particle and Nuclear Study,
		KEK (Tanashi Branch),
		Japan}
	}

\begin{document}
\maketitle

\begin{abstract}
We analyze the invariant mass spectrum of \LamLam\ 
in $\nuc{12}{C}(K^-,K^+\Lambda\Lambda)$ reaction at $P_{K^+}=1.65$ GeV/c
by using a combined framework of IntraNuclear Cascade (INC) model
and the correlation function technique.
The observed enhancement at low-invariant masses 
can be well reproduced with attractive \LamLam\ interactions
with the scattering length either in the range
$a = -6 \sim -4$ fm (no bound state)
or $a = 7 \sim 12$ fm (with bound state).
We also discuss \LamLam\ correlation functions
in central relativistic heavy-ion collisions
as a possible way to eliminate this discrete ambiguity.
\end{abstract}

\section{INTRODUCTION}
\label{sec:Intro}

Extracting hyperon-hyperon interactions is one of the most challenging
current problems in nuclear physics.
It provides us of an opportunity to verify various ideas on
baryon-baryon interactions such as the flavor SU(3) symmetry.
Hyperon-hyperon interactions may also modify
the properties of neutron stars,
in which abundant hyperons are expected to exist.
However,
it is very difficult to determine them experimentally.
For example,
only three double $\Lambda$ nuclei are found in these 35 years,
where only the low-energy ${}^1\hbox{S}_0$ \LamLam\ interaction is accessible.
Therefore, other ways to extract hyperon-hyperon interactions
have been desired so far.

One of these ways is opened up by 
a recent measurement of the invariant mass spectrum of \LamLam\ pair
in $\nuc{12}{C}(K^-,K^+\Lambda\Lambda)$ reaction
by Ahn et al.(KEK-E224 collaboration)~\cite{Ahn}.
The invariant mass spectra, or two-particle momentum correlations,
have been widely used to evaluate the source size when
the final state interactions are negligible or well-known. 
On the other hand, it may possible to extract information of unknown
strong interactions, if we have realistic and reliable source functions.
In this work, we apply the correlation function technique
to the above \LamLam\ invariant mass spectrum, 
and try to determine \LamLam\ interaction at low energies,
by using a source function generated by the IntraNuclear Cascade (INC)
model~\cite{NOHE97},
which well reproduce inclusive $(K^-,K^+)$ spectra.

\begin{figure}[t]
\begin{minipage}[t]{80mm}
\psfig{figure=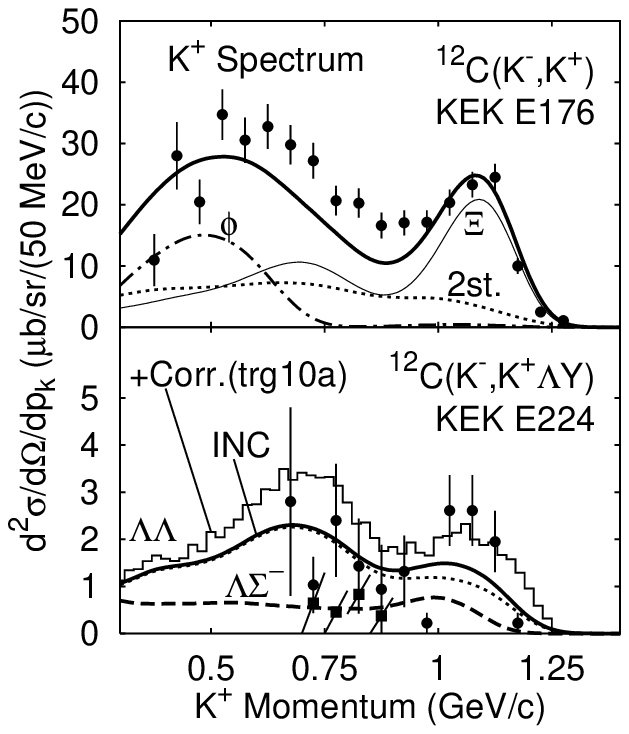,width=70mm}
\vspace*{-10mm}
\caption{%
Calculated momentum spectra of $K^+$ particle
in \nuc{12}{C}$(K^-,K^+)$ (upper panel)
and \nuc{12}{C}$(K^-,K^+\Lambda Y)$ (lower panel) reactions
are compared with data~\protect\cite{Ahn,Iijima,E224}
experiments.
}
\label{fig:kkll}
\end{minipage}
\hspace{\fill}
\begin{minipage}[t]{75mm}
\hspace*{-5mm}
\psfig{figure=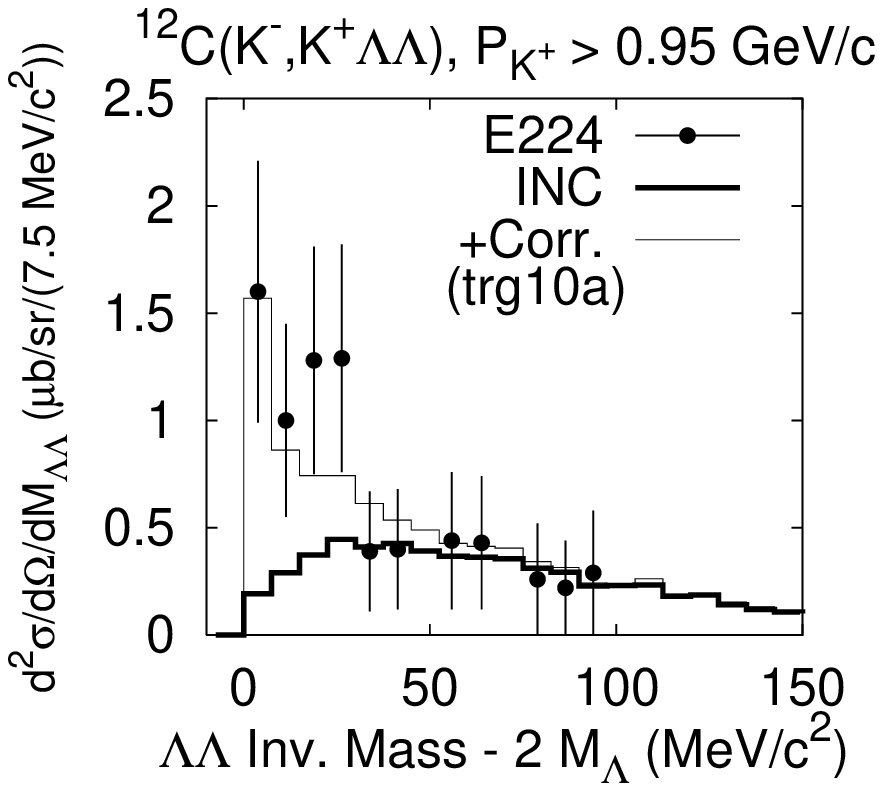,width=80mm}
\vspace*{-15mm}
\caption{%
Calculated $\Lambda$-$\Lambda$ invariant mass spectrum
by using INC model without (thick) and with (thin) correlation effects
are compared with data (solid circle)~\protect\cite{Ahn}.
}
\label{fig:ivll}
\end{minipage}
\end{figure}

\section{INVARIANT MASS SPECTRUM IN \nuc{12}{C}\KK\ REACTION}
\label{sec:KK}
The inclusive spectra~\cite{Iijima} of $(K^-,K^+)$ reaction 
on various targets can be explained well
by the INC model~\cite{NOHE97},
and the predicted reaction mechanism in this model
was experimentally verified recently~\cite{E224}.
In this work, we use this INC model including the baryon mean field potential
effects to generate a hyperon source function.

In Fig.~\ref{fig:kkll},
we compare the calculated $K^+$ momentum spectra
in inclusive $(K^-,K^+)$
and coincidence $(K^-,K^+\Lambda Y)$ reactions
with data~\cite{Ahn,Iijima,E224}.
The calculated results well explains the data
except for the underestimate of two lambda production
at around $P_{K^+}\simeq 1.1$ GeV/c.
This underestimate is concentrated in the low invariant mass region of
\LamLam, as shown in Fig.~\ref{fig:ivll}.
This is as expected,
since the correlation caused by the final state interaction 
modify the invariant mass spectrum
at low mass region most effectively.

The enhancement of the two-particle momentum spectrum has been
extensively studied in the context of correlation function.
For example, 
in the correlation function formula~\cite{HBT},
the probability to find particle pair at momenta $\vec{p}_1$ and $\vec{p}_2$
reads, 
\beq
\label{eq:correlation}
P(\vec{p}_1,\vec{p}_2)
	= \int d^4x_1\,d^4x_2\, S_{12} (\vec{p}_1,x_1,\vec{p}_2,x_2)\ 
		\left| \psi^{(-)}(\vec{r}_{12};\vec{k}) \right|^2\ ,
\eeq
where $\vec{r}_{12}$ is the relative distance
of two particles at particle creation, 
$S_{12}$ denotes the source function,
and the wave function $\psi^{(-)}$ is chosen
to have the outgoing relative momentum
$\vec{k} = (\vec{p}_1 - \vec{p}_2) / 2\hbar$.
%
%
In this work, we limit ourselves to working in a single channel description
of $\Lambda\Lambda$,
and assume that two $\Lambda$ particles are in their spin singlet state.
Then, the relative wave function of \LamLam\ 
at small relative momentum is simplified as follows.
\beq
\label{eq:llwf}
\psi^{(-)}(\vec{r};\vec{k})
\simeq
	\sqrt{2} \left[
		\cos(kr\cos\theta)-j_0(kr)+e^{-i\delta_0} u_0(r;k)
	\right] \ .
\eeq

\begin{table}[t]
\caption{
	Two-range Gauss (trg) \LamLam\ potential parameters,
	in the unit of fm (MeV) for $\mu_l, a$, and $\reff$ 
	($v_l, v_s$, and binding energies).
	The shorter range $\mu_s=0.45$ fm is fixed for simplicity.
}
\label{table:llpars}
\begin{tabular}{r|crrrrr|crrrrrr}
\hline
\hline
$\mu_l$ &
	&$v_l$	&$v_s$	
	& $a$	&\reff	
	&$\tilde{\chi}^2$
	&
	&$v_l$	&$v_s$	
	& $a$	&\reff	
	&$\tilde{\chi}^2$
	& B.E. \\
\hline
0.6	&06a	& -900	&1440	& -4.4	& 1.6	& 0.34
	&06b	& -950	&1310	&  7.5	& 1.2	& 0.37	& 0.72\\
0.8	&08a	& -230	& 470	& -5.0	& 1.8	& 0.36
	&08b	& -270	& 410	&  8.5	& 1.3	& 0.40	& 0.56\\
1.0	&10a	& -105	& 200	& -6.2	& 2.0	& 0.39
	&10b	& -135	& 210	& 11.5	& 1.6	& 0.43	& 0.29\\
\hline
\hline
\end{tabular}
\end{table}

%
Based on these formula,
we have calculated the \LamLam\ invariant mass spectrum
including the correlation effects.
%
We have used here phenomenological \LamLam\ potentials
parametrized in a two-range Gaussian form, 
$V_\sLamLam(r) = v_l \exp(-r^2/\mu_l^2) + v_s \exp(-r^2/\mu_s^2)$.
%
As shown by the thin histogram in Fig.~\ref{fig:ivll},
the enhancement at low invariant masses can be reproduced
with an appropriate potential,
although the second "peak" around 25 MeV is difficult to explain.
At higher invariant masses 
the correlation effects are washed out 
by the oscillation of $|\psi|^2$
in the source spreading upto around $r_{12} \simeq$ 5 fm. 

We have searched for the $\chi^2$ local minima in the $(v_l, v_s)$ plane
for a given longer range parameter,
$\mu_l = 0.6 \sim 1.0$ fm (Table~\ref{table:llpars}).
%
%
In Fig.~\ref{fig:Nmg-ar}, we compare the extracted \LamLam\ potentials
with the Nijmegen models.
From this comparison,
model ND (NF) with the hardcore radius $r_c \simeq 0.50 (0.46) $ fm,
and model NSC89 with the cutoff mass $M_{cut} \simeq 920$ MeV
are expected to give reasonable enhancement
in the \LamLam\ invariant mass spectrum.

\section{DO TWO LAMBDAS BOUND ?}

As shown in Table~\ref{table:llpars}, 
all the best fit parameters (trg-a) give negative scattering length
($\delta_0 \sim -ak$),
implying that there is no bound state.
However,
there is another local minimum (trg-b)
in the region where there is a bound state, $a > 0$.
This double-well structure appears because the enhancement
is described by the wave function squared.
With the long wave approximation,
the wave function becomes essentially a constant, 
then the correlation effect can be factorized as follows~\cite{SAT89},
\begin{equation}
P(\vec{p}_1,\vec{p}_2)
        \simeq\ 
	2 F(k)\ P_c(\vec{p}_1,\vec{p}_2)\ ,\quad
P_c = \int d^4x_1\,d^4x_2\, S_{12}\ , \quad
F(k) = (1-a/b)^2 - ck^2,
\end{equation}
where $b$ denotes the intrinsic range.
There are two solutions of $a$ giving the same enhancement at low energies, 
$a \sim b\ \left( 1 \pm \sqrt{F(0)} \right)$.
When the spectrum is enhanced at low energies, $F(0) > 1$,
one of them is negative and the other becomes positive.

\begin{figure}[t]
\begin{minipage}[t]{80mm}
\psfig{figure=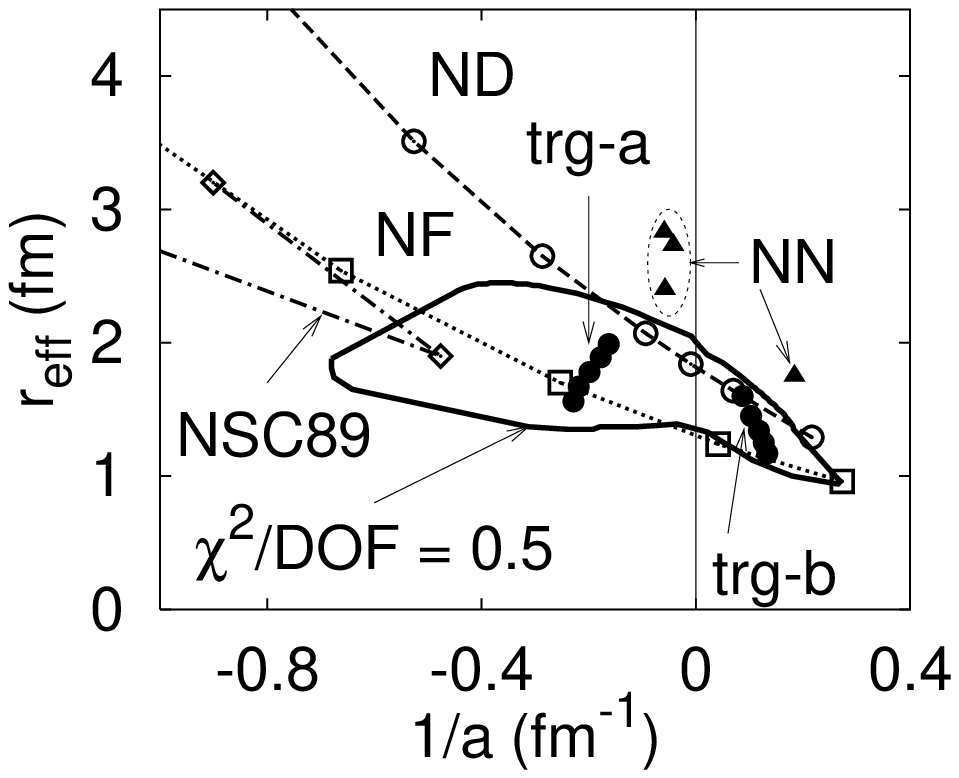,width=70mm}
\vspace*{-10mm}
\caption{%
Extracted $\Lambda$-$\Lambda$ scattering parameters.
Fitted two-range gaussian potentials (solid circles)
are compared with Nijmegen models (open marks).
Inside the thick solid line,
potential with $\tilde{\chi}^2 \leq$ 0.5 
exists in the range $0.6 \leq \mu_l \leq 1.0$ fm.
}
\label{fig:Nmg-ar}
\end{minipage}
\hspace{\fill}
\begin{minipage}[t]{75mm}
\hspace*{-5mm}
\psfig{figure=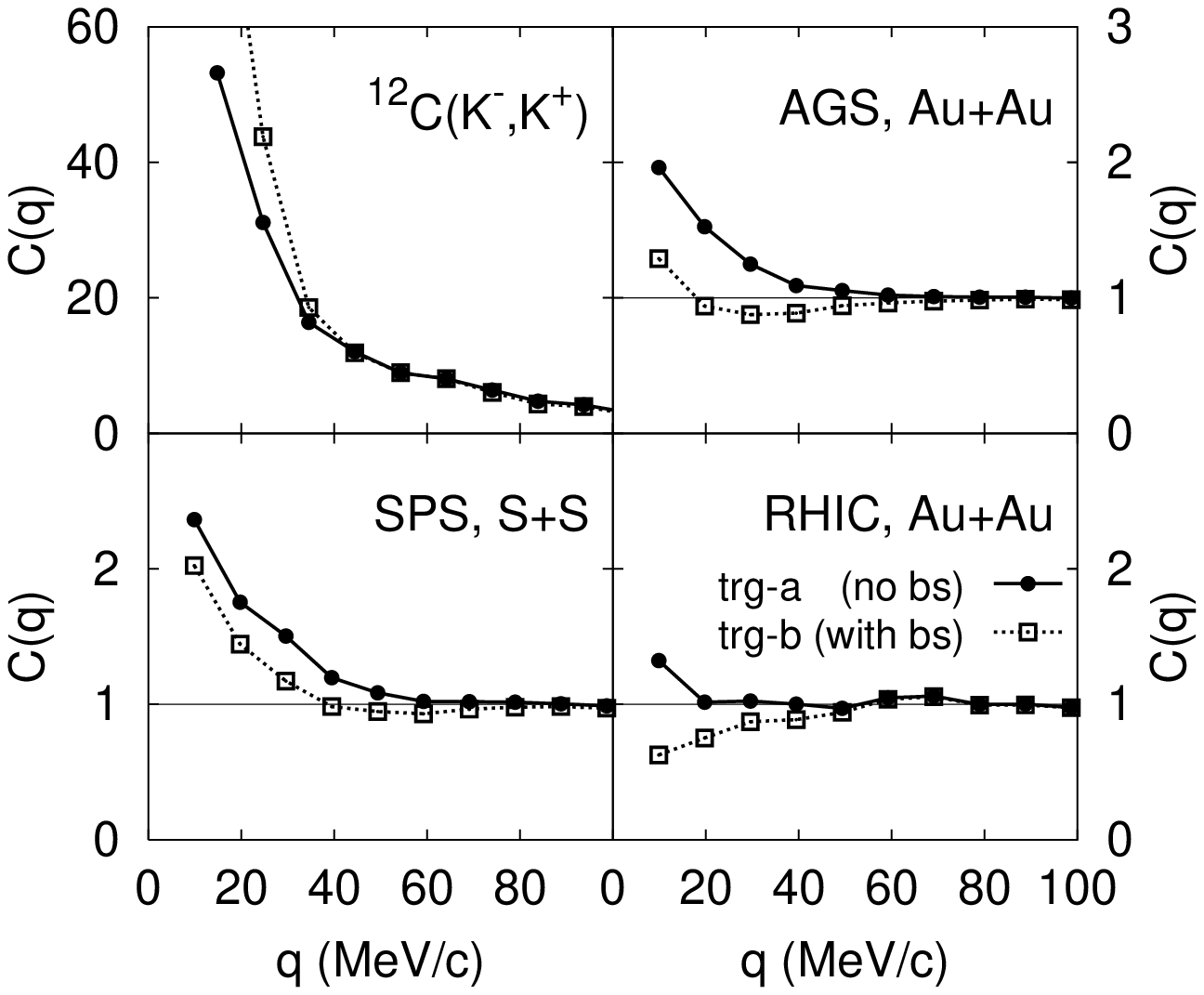,width=80mm}
\vspace*{-15mm}
\caption{%
$\Lambda$-$\Lambda$ correlation function in $(K^-,K^+)$ reaction
and central relativistic heavy-ion collisions at AGS, SPS, and RHIC energies.
The results with trg10a (solid circles) and trg10b (open squares)
are compared.
}
\label{fig:HIC-corr}
\end{minipage}
\end{figure}

In order to overcome this problem and to distinguish these two solutions,
it may be helpful to use relativistic heavy-ion collisions,
where the source size is much larger than that in $(K^-,K^+)$ reaction
and the long wave approximation would not work.
For example, 
the scattering wave function have at least
one node in the range $r \leq a$ when a bound state exists,
which suppresses the contribution at around $r \sim a$
in two-particle correlations.

In Fig.~\ref{fig:HIC-corr}, 
we show the calculated correlation function 
in relativistic heavy-ion collisions
with the source function generated by using a recently developed
cascade model, JAM~\cite{JAM}.
As expected in the above discussion,
two-particle correlations are suppressed at small momenta with trg10b 
(with bound state) compared with trg10a (no bound state).

\section{Summary}

We have analyzed the \LamLam\ invariant mass spectrum
in $\nuc{12}{C}(K^-,K^+\Lambda\Lambda)$ reaction
by using the classical source function given by IntraNuclear Cascade (INC)
model combined with the correlation function formula,
which takes account of \LamLam\ interaction.
Within a single channel description of \LamLam, and under the assumption
that two lambdas are in their spin-singlet state,
we limit the allowed range of \LamLam\ scattering parameters.
We have also discussed a possibility to use relativistic heavy-ion
collisions as one of the ways to distinguish the \LamLam\ potentials,
which reproduce the invariant mass spectrum of \LamLam\ equally well
in $(K^-,K^+)$ but have different sign of the scattering length.

\medskip

The authors would like to thank
Dr. J. K. Ahn and Prof. J. Randrup for fruitful discussions.
This work was supported in part by
the Grant-in-Aid for Scientific Research
(Nos.\	07640365,	
	08239104	
and	09640329)	
from the Ministry of Education, Science and Culture, Japan,

%
%
%
%
%
%


\begin{thebibliography}{99}
\bibitem{Ahn}
J.K. Ahn et al. (KEK-E224 collaboration),
	Nucl. Phys. A639 (1998) 379c; 
	Phys. Rev. Lett., in press.

\bibitem{NOHE97}
Y. Nara, A. Ohnishi, T. Harada, and A. Engel,
	Nucl. Phys. A614 (1997) 433.

\bibitem{Iijima}
	T. Iijima {\it et al}. (KEK-E176 collaboration),
		Nucl. Phys. A546 (1992) 588.

\bibitem{E224}
J.K. Ahn et al. (KEK-E224 collaboration),
	Nucl. Phys. A625 (1997) 231.


%

\bibitem{HBT}
W. Bauer  et al.,
	Annu. Rev. Nucl. Part. Sci. 42 (1992) 77,
	and references therein.

\bibitem{SAT89}
I. Slaus, Y. Akaishi and H. Tanaka,
	Phys. Rep. 173 (1989) 257.

\bibitem{JAM}
	Y. Nara, Nucl. Phys. A638 (1998) 555c; Eprint nucl-th/9802016.

\end{thebibliography}
\end{document}